\documentclass{optica-article}
    \usepackage{mathtools}
\journal{opticajournal}

\articletype{Research Article}

\usepackage{lineno}

\begin{document}

\title{Ultrawide Edge State Supercontinuum in a Floquet-Lieb Topological Photonic Insulator}

\author{Hanfa Song\authormark{1*}, Tyler Zimmerling\authormark{1}, Bo Leng \authormark{1} and Vien Van\authormark{1}}

\address{\authormark{1} Department of Electrical and Computer Engineering, University of Alberta, Edmonton, Alberta, Canada }

\email{\authormark{*}hanfa@ualberta.ca} 



\begin{abstract}
Conventional topological photonic insulators typically have narrow nontrivial band gaps truncated by broad dispersive bulk bands, resulting in limited edge mode transmission bandwidths that can be exploited for potential applications.  Here we propose and demonstrate the first Floquet-Lieb topological photonic insulator with all flat bands which can support continuous edge mode transmission across multiple Floquet-Brillouin zones.  
This supercontinuum of edge states results from the orthogonality between the flat-band modes and the edge states, allowing for continuous excitation of the latter without scattering into the bulk modes.
Moreover, we show that these flat bands are perfectly immune to random variations in the on-site potential, regardless of how large the perturbations are, thus ensuring complete robustness of the edge modes to this type of disorder.
We realized Floquet-Lieb insulators using 2D microring resonator lattices with perfect nearest-neighbor couplings.
Transmission measurements and direct imaging of the scattered light distributions showed an edge mode supercontinuum spanning more than three microring free spectral ranges.
These broad edge mode transmission bands persisted even in the presence of lattice disorder caused by fabrication imperfection.  The proposed Floquet-Lieb insulator can potentially be used to realize topological photonic devices with ultrawide bandwidths and super robustness for applications in integrated quantum photonics and programmable photonic circuits.
\end{abstract}


\section{Introduction}
Topological photonic insulators (TPIs) are photonic lattices which can host edge states in topologically nontrivial band gaps \cite{lu2014topological,ozawa2019topological,khanikaev2017two,tang2022topological}. 
These structures have attracted much attention since they provide a versatile platform for probing various quantum  phenomena in solid state systems\cite{yan2021quantum,saxena2022photonic,reitz2022cooperative,von202040}.  In addition, the topological protection property of the edge states can potentially be exploited to realize robust photonic devices that are resistant to disorders, such as those caused by fabrication imperfection \cite{shalaev2019robust,hafezi2011robust,song2022analytical}.  Although TPIs have been realized in various 2D and 3D systems, the nontrivial band gaps hosting edge modes typically exist only over a narrow frequency range, which limits their potential applications \cite{yang2019realization,hafezi2011robust}.  
In particular, while Floquet TPIs based on periodically-driven systems can exhibit richer topological behavior than undriven systems, their edge mode spectra cannot exceed a Floquet-Brillouin (FB) zone due to the presence of dispersive bulk bands.
Moreover, these bulk bands make the lattice less resistant to disorder, since large perturbations exceeding the gap energy could push a gap state into the bulk bands, thus destroying the edge mode \cite{zhang2021superior,harari2018topological}.  Ideally, the most robust topological system that can be created would have all nontrivial band gaps separated by flat bulk bands that remain flat even in the presence of disorder.  Here we show that such a system could be realized by a Floquet-Lieb insulator (FLI), which is a Floquet topological insulator with all flat bands. Moreover, these flat-band modes can be suppressed to provide a supercontinuum of edge modes spanning multiple FB zones.

Flat-band systems are characterized by completely flat or dispersionless energy bands extending over the whole Brillouin zone \cite{harris1997geometrical,leykam2018artificial,mukherjee2015observation,shen2010single}. 
The zero group velocity and strong localization of the flat-band modes make these systems ideal for probing various phenomena such as frustrated hopping \cite{bergman2008band,baboux2016bosonic}, Anderson localization \cite{chalker2010anderson,vakulchyk2017anderson} and fractional quantum Hall effect \cite{tang2011high,yang2012quantum}.  Flat-band systems with various lattice geometries have been proposed and demonstrated in solid state systems and optical lattices \cite{hyrkas2013many,huber2010bose,green2010isolated,weeks2010topological,leykam2018artificial,ivanov2020edge}, among which the 2D Lieb lattice stands out due to its simple topology and unique properties. Notably, due to the particle-hole symmetry of the system, the flat band of a Lieb lattice is insensitive to disorder in the coupling strengths \cite{mukherjee2015observation}. 
In the photonics domain, Lieb lattices have been realized using 2D arrays of uniformly-coupled waveguides, with the flat-band modes shown to support non-diffracting light propagation \cite{mukherjee2015observation,guzman2014experimental,xia2018unconventional}. However, these systems are gapless so they do not support edge modes.  Flat-band states have also been reported in a Floquet TPI based on periodically-coupled waveguide array \cite{maczewsky2017observation}, although this system is not a Lieb insulator and the transmission spectrum of edge modes was not investigated.
 We note that coupled microring arrays in the Lieb lattice configuration have been used to realize anomalous Floquet insulators \cite{zimmerling2022broadband,dai2022topologically}, but these systems were not designed to have all flat bands since the couplings between neighbor resonators were not perfect.  As a result the edge modes existed only over a small fraction of the FB zone due to the appearance of dispersive bulk bands.


In this paper we propose and experimentally demonstrate the first Floquet-Lieb TPI with all flat bands which can support a continuous edge mode spectrum exceeding a FB zone. Our FLI is based on a 2D Lieb lattice of microring resonators with perfect nearest-neighbor couplings.    
Owing to the periodic circulation of light in the microrings, the lattice emulates a periodically-driven system with quasi-energy bands repeating every microring free spectral range (FSR).  
By exploiting the orthogonality between the flat-band modes and edge modes, the latter can be excited continuously across the flat band energies without coupling into these modes, resulting in an edge state supercontinuum that in principle can extend indefinitely.  Furthermore, we show that these edge modes are perfectly immune to variations in the microring resonance frequencies, or on-site potentials, regardless of how large the frequency detunes are. 
To our knowledge this remarkable robustness behavior has not been shown in any other topological insulator.
We note that the crucial difference between our FLI and previous realizations of microring Lieb lattices in \cite{zimmerling2022broadband,dai2022topologically} is the perfect coupling condition, which gives rise to the wide edge mode supercontinuum and perfect immunity to resonance frequency disorder.
We realized FLIs on a silicon photonic platform, achieving an ultrawide edge mode transmission spectrum of over 20 nm, equivalent to more than three FSRs, in the region of near perfect coupling.  
We also verified the existence of the edge modes across the entire transmission band by near-infrared (NIR) imaging of the scattered light intensity distributions.
In addition, measurement of a random sample of fabricated lattices showed that these wide transmission bands persist in the presence of random variations due to fabrication, thus confirming the robustness of the edge modes.
The ultrawide edge state spectrum of the FLI, combined with its realization on an integrated platform, potentially enables topological photonic devices with broad bandwidths and superior robustness for applications in integrated quantum photonics and programmable photonic circuits \cite{song2022robust,zand2020effects}.


\section{Floquet-Lieb Insulator based on 2D coupled microring lattice}

\begin{figure}[t]
\centering\includegraphics[width=0.5\linewidth]{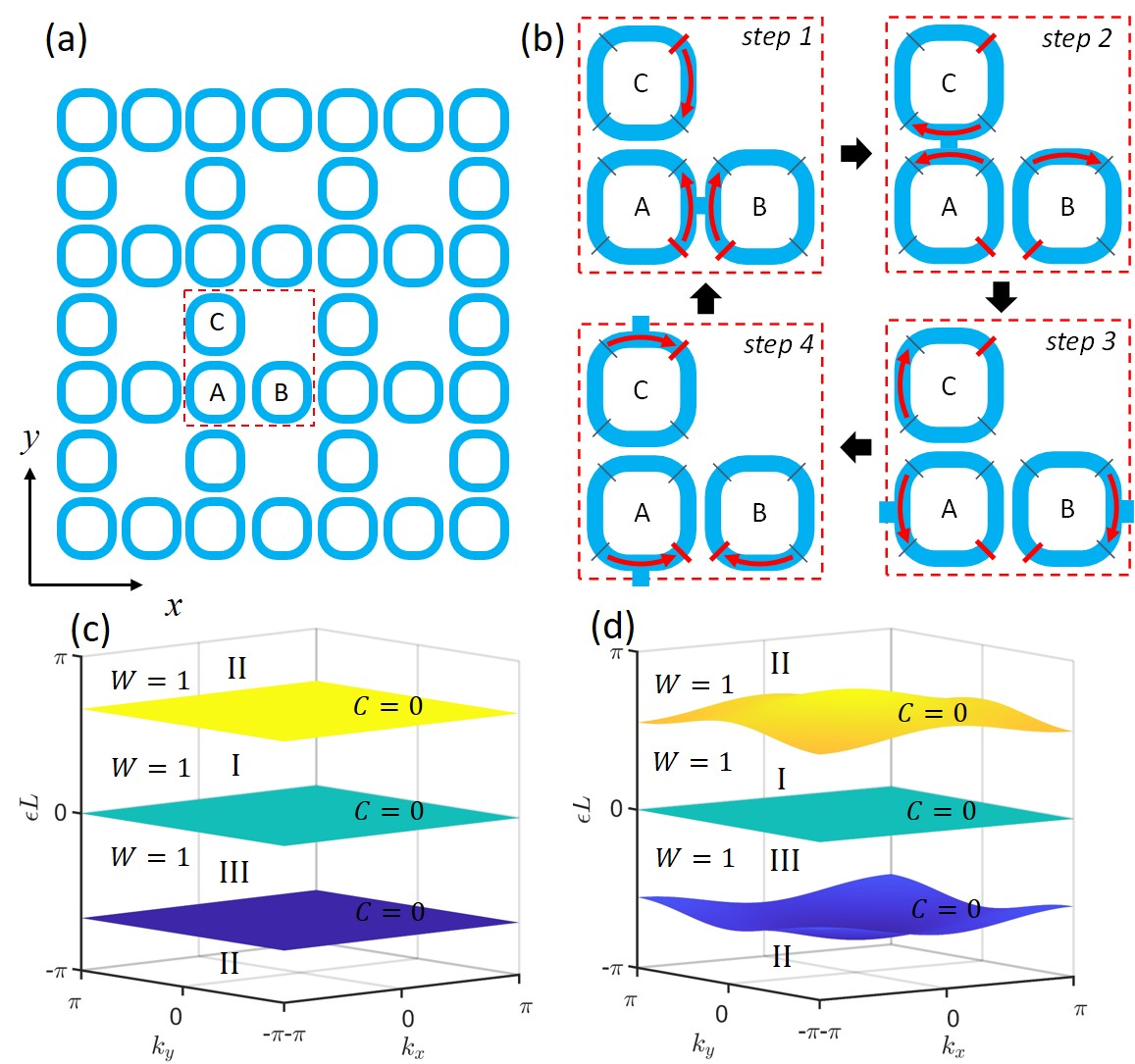}
 \caption{(a) Schematic diagram of a 2D FLI microring lattice. Each unit cell consists of 3 identical microrings $A$, $B$ and $C$ with identical nearest-neighbor coupling angle $\theta$. (b) Sequence of four coupling steps describing the evolution of light over one period in each unit cell. (c) and (d) Quasi-energy band diagrams of a FLI with (c) perfect coupling ($\theta=0.5\pi$) and (d) imperfect coupling ($\theta=0.45\pi$). The Chern numbers ($C$) of the bands and winding numbers ($W$) of the gaps are also indicated.}
\label{fig:FLI_schematic}
\end{figure}

Figure \ref{fig:FLI_schematic}(a) shows a schematic diagram of the FLI based on a 2D Lieb lattice of coupled microring resonators, with each unit cell consisting of three identical microrings, $A$, $B$ and $C$. Each microring is assumed to support only a clockwise or counterclockwise propagating mode, with negligible scattering into the reverse mode. We also assume identical coupling between adjacent microrings, with the coupling strength denoted by a coupling angle $\theta$ defined such that the fraction of optical power transfer between the two rings is $\sin^2{\theta}$. As light circulates around each microring, it is periodically coupled to neighbor rings with a period equal to the microring circumference $L$. The lattice thus emulates a Floquet system with the direction of light propagation $(z)$ along the microring waveguides taking the role of time \cite{afzal2018topological}.  Specifically, starting from the points indicated by the red bar on each microring in Fig. \ref{fig:FLI_schematic}(b) and tracing the path of light propagation around the rings, the evolution of the system over each period can be decomposed into a sequence of four coupling steps as shown in the figure. 
Using waveguide coupled mode theory, we can express the evolution of a Bloch state $|\psi\rangle = [\psi_A, \psi_B, \psi_C]^T$ along the direction of light propagation $z$ around the microrings as (see Appendix A for derivation) \cite{afzal2018topological}
\begin{equation}
-i \frac{\partial}{\partial z}|\psi(\textbf{k},z)\rangle= [ \beta I + H_{FB}(\textbf{k},z) ]|\psi(\textbf{k},z)\rangle
\label{CMT}
\end{equation}
where $\beta$ is the propagation constant of the microring waveguides, $\textbf{k} = (k_x, k_y)$ is the crystal momentum vector, and $I$ is the $3\times3$ identity matrix.  The Floquet-Bloch Hamiltonian $H_{FB}$ describes the periodic coupling sequence and is given by the Hamiltonian $H_j$ ($j = 1\dots4$) in coupling step $j$ (see Appendix A).
Suppressing the dynamic phase term $e^{i \beta L}$, the evolution of the system over one period is described by the Floquet operator $U_F(\textbf{k})=\mathcal{P} e^{i\int_0^L{H_{\mathrm{FB}} dz}}$, where $\mathcal{P}$ is the path ordering operator.  
From the Floquet operator, we can define an effective Hamiltonian $H_{\mathrm{eff}}(\textbf{k}) = -(i/L) \ln{U_F(\textbf{k})}$, which captures the stroboscopic motion of the system over one period.  In this work we are most interested in the FLI with perfect coupling ($\theta = \pi/2$), which corresponds to 100$\%$ power transfer between neighbor rings in each coupling step.  For this lattice the effective Hamiltonian can be explicitly computed to give
\begin{equation}
      H_{\mathrm{eff}}(\textbf{k})= \frac{2\pi/L}{3\sqrt{3}}
  \begin{bmatrix}
         0 &  -e^{-ik_y} &  -e^{i(k_x-k_y)} \\
     -e^{ik_y}& 0 & ie^{ik_x}\\
      -e^{-i(k_x-k_y)} & -ie^{-ik_x} & 0\\
  \end{bmatrix}.
  \label{Heff}
\end{equation}

The quasi-energy bands of the FLI are obtained from the spectrum of the Floquet operator,
$U_{F}(\textbf{k})\left|\Phi_n(\textbf{k})\right\rangle=e^{i \varepsilon_n(\textbf{k}) L}\left|\Phi_n(\textbf{k})\right\rangle$,
where $\varepsilon_n(\textbf{k})$ is the $n$-th quasi-energy band and $|\Phi_n\rangle$ the associated Floquet-Bloch eigenstate.
Unlike non-driven systems, the quasi-energy bands of the Floquet insulator repeat with a periodicity of $2\pi/L$, which corresponds to the microring's FSR.  Each FSR or FB zone has a flat band $\varepsilon_0(\textbf{k}) = 0$ and upper and lower dispersive bands $\varepsilon_{\pm}$ given by
\begin{equation}
    \varepsilon_{\pm}(\textbf{k})L = \pm \{\pi - \cos^{-1}[\kappa^4/2 - \tau^4 + \kappa^2 \tau (\cos k_x + \cos k_y)]\},
    \label{dispersive_bands}
\end{equation}
where $\kappa = \sin \theta$ and $\tau = \cos \theta$.  For the FLI with perfect coupling we obtain three flat bands at $\varepsilon L=0,\pm{2\pi/3}$ in each FSR.  As the coupling angle decreases from $\pi/2$, the upper and lower bands $\varepsilon_{\pm}$ become broader and more dispersive but the zero-energy band $\varepsilon_0$ is always flat. Figs. \ref{fig:FLI_schematic}(c) and (d) show the quasi-energy band diagrams of FLI lattices with perfect coupling and imperfect coupling ($\theta=0.45\pi$), respectively, with both supporting three open band gaps (labeled I, II and III) in each FSR.  From Eq.(\ref{dispersive_bands}) we find that the three gaps stay open for all values of the coupling angle except for $\theta \sim 0.36\pi$, when gap II closes. This is a key difference between our FLI and undriven Lieb lattices \cite{xia2018unconventional,guzman2014experimental,mukherjee2015observation}, which are gapless since the three bands become degenerate at $(k_x, k_y) = (\pm \pi/2, \pm \pi/2)$.  While these degeneracies can be lifted by introducing next-nearest neighbor coupling between sites $B$ and $C$ \cite{bhattacharya2019flat,yarmohammadi2018controllable}, these couplings are generally difficult to realize physically. For our FLI lattice, even though there is no direct coupling between rings $B$ and $C$, the next-nearest neighbor coupling terms ($\pm i e^{\pm i k_x}$) naturally appear in the effective Hamiltonian $H_{\mathrm{eff}}$ in Eq.(\ref{Heff}) due to the non-commutativity of the Hamiltonians $H_j$ of the coupling steps \cite{tsay2011analytic}. As a result, gaps I and III are always open and can thus host edge states.
\begin{figure}[t]
\centering\includegraphics[width=0.8\linewidth]{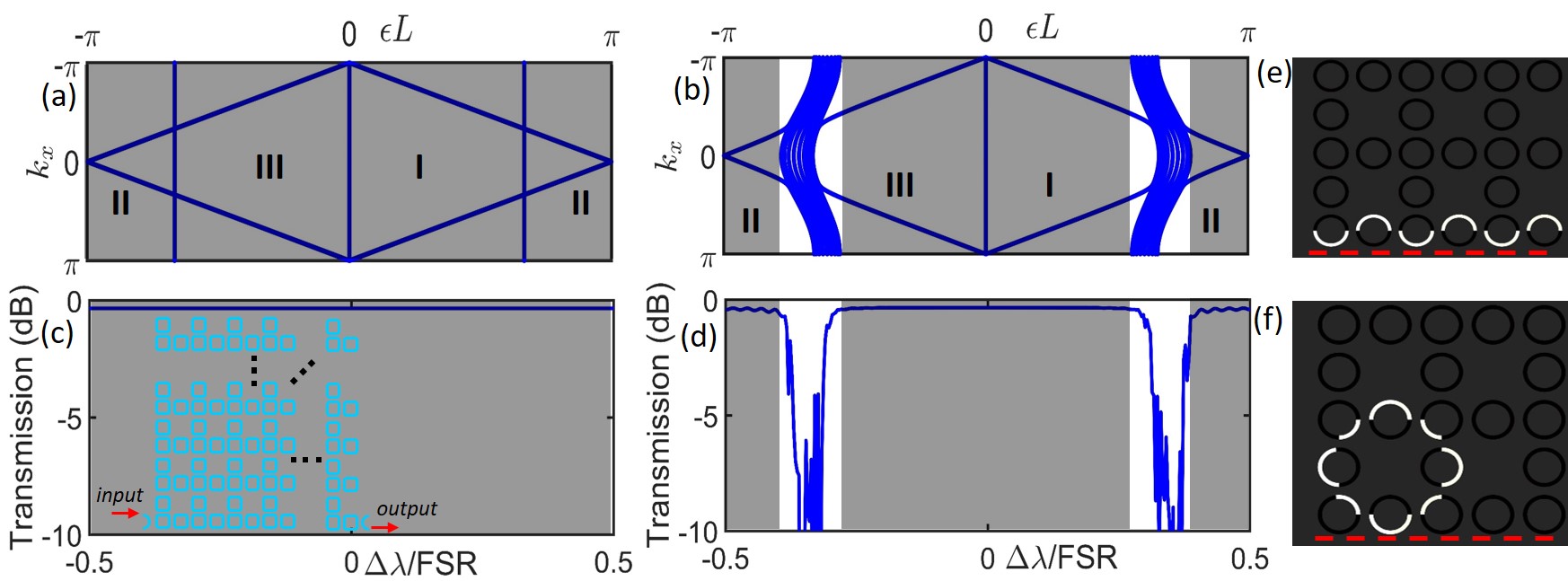}
 \caption{(a), (b) Projected band diagram of a semi-infinite FLI with coupling angle (a) $\theta=0.5\pi$ and (b) $\theta=0.45\pi$. Grey areas indicate the band gaps hosting edge modes. (c), (d) Simulated edge mode transmission spectrum over one microring FSR of a $10 \times 10$ unit cell lattice (shown in inset of (c)) with (c) $\theta=0.5\pi$ and (d) $\theta=0.45\pi$. (e), (f) Spatial intensity distributions of (e) forward-propagating edge mode and (f) flat-band mode at zero quasi-energy.}
\label{fig:Semi_edgemode}
\end{figure}

\begin{figure}[t]
\centering\includegraphics[width=0.8\linewidth]{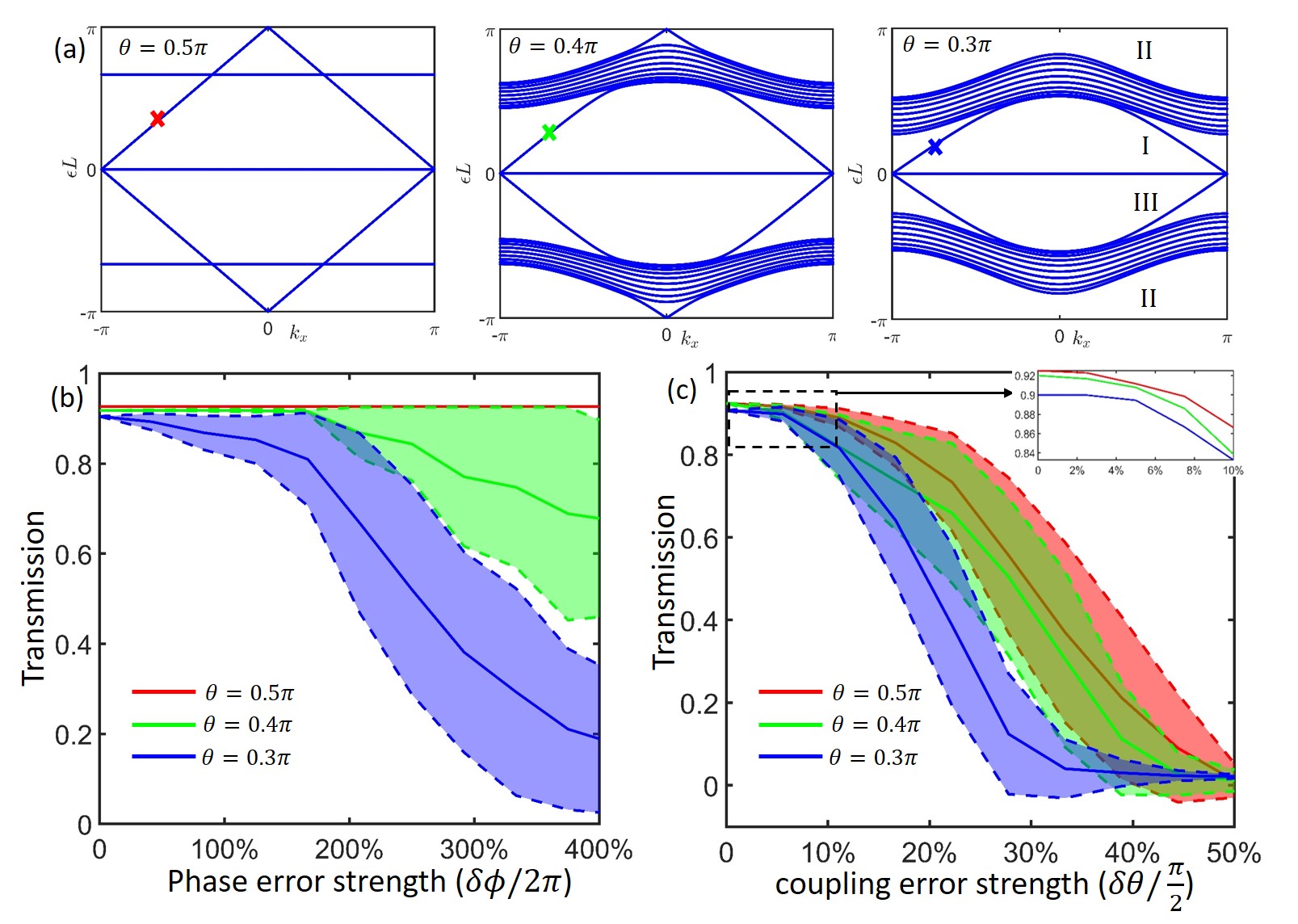}
 \caption{(a) Projected band diagrams of FLI lattices with coupling angle $\theta = 0.5\pi, 0.4\pi$, and $0.3\pi$. Cross marks indicate the energy of the excited edge mode.  (b), (c) Edge mode transmission of the 3 FLI lattices in the presence of uniformly-distributed variations in (b) microring roundtrip phases and (c) coupling angles. Solid lines and shaded regions indicate the transmission mean and standard deviation, respectively.}
\label{fig:Robustness}
\end{figure}

The FLI microring lattice can be classified topologically by a winding number associated with each band gap.  In particular, it was shown in \cite{afzal2018topological} that for $\theta \gtrsim 0.36\pi$, all three gaps have nontrivial winding numbers even though the Chern number of each bulk band is trivial.  The lattice thus behaves as an anomalous Floquet insulator, which supports topologically protected edge states in all three band gaps. 
Figures \ref{fig:Semi_edgemode}(a) and (b) show the projected band diagrams of a FLI lattice with 10 unit cells in the $y$ direction and infinite extent in $x$ for the cases of perfect and imperfect coupling, respectively.
Each band gap is seen to support two counter-propagating edge states.  For the lattice with perfect coupling, the edge modes continuously span across the whole FSR except at the points of intersection with the flat bands.  
Figures \ref{fig:Semi_edgemode}(e) and (f) show the spatial field distributions of the forward-propagating edge mode and a flat-band mode (at zero quasi-energy), respectively.  The edge mode is seen to be localized along the bottom lattice boundary while the flat-band mode is strongly confined in a resonant loop pattern.  As shown in the Appendix C, these two modes are orthogonal so they do not couple at the crossing points.  As a result, a continuous spectrum of edge modes can be excited over the entire FB zone without scattering into the flat-band modes.  To verify this, we simulated the transmission of an edge mode along the bottom boundary of a finite lattice with 10 $\times$ 10 unit cells, as shown in the inset of Fig. \ref{fig:Semi_edgemode}(c). We used the same microring size as in the fabricated device in Fig. \ref{fig:SEMnlongscan} (a), with waveguide group index $n_g = 4$ and propagation loss of 3 dB/cm around the 1550 nm wavelength.
The edge mode was excited at the bottom left corner of the lattice through an input waveguide and the transmitted power was monitored at the bottom right corner through an output waveguide (see Appendix B for details of the simulation method).
For the case of perfect coupling, we obtained a constant edge mode transmission spectrum across the entire microring FSR, as shown in Fig. \ref{fig:Semi_edgemode}(c), indicating that the flat-band modes are not excited at the frequencies of degeneracy.
On the other hand,  when the coupling is imperfect ($\theta = 0.45\pi)$, bulk modes in the dispersive bands at quasi-energies $\varepsilon_{\pm}$ are excited, resulting in two broad dips in the transmission spectrum, as shown in Fig. \ref{fig:Semi_edgemode}(d). However, the flat-band mode at zero quasi-energy remains unexcited even for imperfect coupling, so that a wide edge mode spectrum can still be obtained around the microring resonance frequency.  
\par
The FLI lattice with perfect coupling not only provides the broadest edge mode continuum possible
but the edge states are also completely immune to random variations in the microring resonant frequencies.  This remarkable behavior stems from the fact that the flat bands of the effective Hamiltonian in Eq.(\ref{Heff}) are insensitive to disorder in the diagonal elements, implying that in the presence of random microring resonant frequency detunes, the bands may be shifted in energy but they always remain flat (see proof in Appendix D).  This is in contrast to a conventional Lieb lattice, whose flat band becomes dispersive in the presence of diagonal disorder \cite{mukherjee2015observation}. 
In addition, the flat-band modes of the FLI remain orthogonal to the edge states so that there is no coupling between the two modes, regardless of how much the flat bands are shifted due to perturbation.
To verify this super robustness behavior, we performed simulations of edge mode propagation along the bottom boundary of a FLI lattice with $10\times10$ unit cells in the presence of uniformly-distributed variations in the microring roundtrip phases. For comparison, we considered three FLI lattices with coupling angle $\theta=0.5\pi, 0.4\pi$ and $0.3\pi$, the last case corresponding to a conventional Chern insulator (CI) with a trivial gap II \cite{afzal2018topological}.  The band diagram of each lattice is shown in Fig. \ref{fig:Robustness}(a).
For each lattice we excited an edge mode at the center of nontrivial gap I, as indicated by the cross marks in Fig. \ref{fig:Robustness}(a).  Figure \ref{fig:Robustness}(b) shows the edge mode transmission as a function of the maximum phase error bound, with the transmission mean and standard deviation indicated by the solid lines and shaded areas, respectively.  We observe that for the FLI with perfect coupling, the edge mode transmission is completely insensitive to the phase error strength, implying perfect immunity to random phase errors. This behavior can be intuitively understood from the fact that for perfect coupling, the rings no longer act as resonators so the edge mode transmission does not depend on the roundtrip phases.  When the coupling is imperfect, the transmission degrades for large phase errors due to the appearance of dispersive bulk bands, although the FLI with all nontrivial gaps ($\theta = 0.4\pi$) performs better than the CI lattice.  This is because as the phase error increases, the nontrivial gaps (I and III) of the CI close so that no edge state can exist.  
We note that for random variations in the coupling angle (disorder in the off-diagonal elements of the effective Hamiltonian), all three bands of the FLI with perfect coupling still remain flat; however, these modes are no longer orthogonal to the edge states at the crossing frequencies.
As a result, the edge modes do not have perfect immunity to coupling disorder but show degradation in transmission due to scattering into the flat bands as the coupling error increases, as shown in Fig. \ref{fig:Robustness}(c).  Nevertheless, the FLI with perfect coupling still exhibits superior robustness to the lattices with imperfect coupling because it is less likely for the edge state to exactly coincide in energy with a randomly-shifted flat band than a dispersive bulk band.

\section{Realization of Floquet-Lieb microring lattice on a silicon photonic platform}
\begin{figure}[t]
\centering\includegraphics[width=0.8\linewidth]{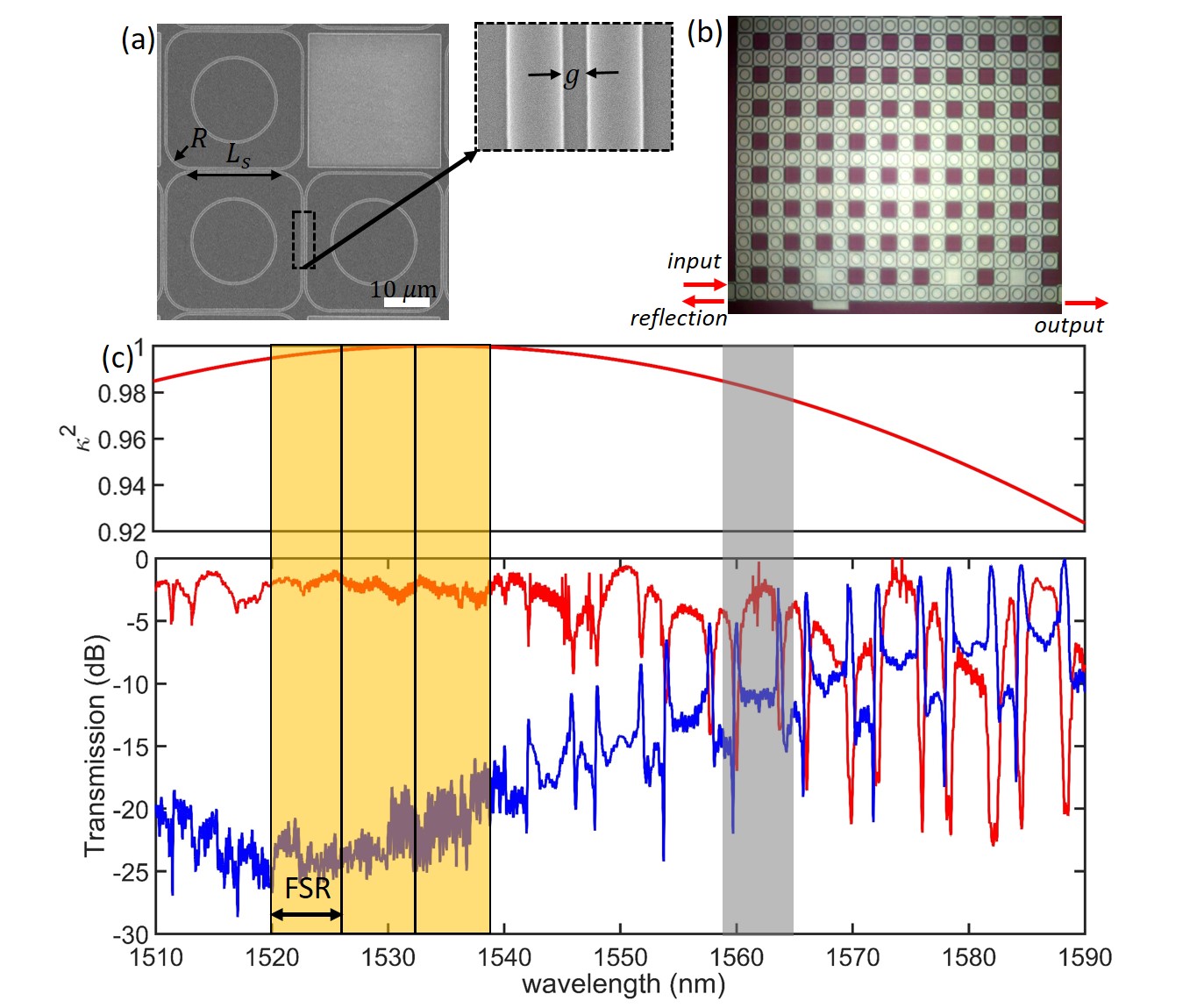}
 \caption{(a) Scanning electron microscope image of a unit cell of the FLI microring lattice. (b) Optical microscope image of the FLI lattice with 10 $\times$ 10 unit cells showing the locations of the input port, output port and reflection port. (c) Lower panel shows the transmission spectra measured at the output port (red) and reflection port (blue) of the FLI lattice. 
 Upper panel shows the simulated wavelength dispersion of the power coupling coefficient of the coupler between adjacent microrings.}
\label{fig:SEMnlongscan}
\end{figure}
We fabricated FLI lattices with $10 \times 10$ unit cells on a silicon-on-insulator substrate using the AMF silicon photonics foundry \cite{AMF}. The microring rib waveguides were designed to support the fundamental transverse electric (TE) mode with 500 nm core width, 130 nm rib height, and 90 nm slab thickness on a 2 $\mu$m-thick $\text{SiO}_2$ substrate.  We used square-shaped ring resonators with straight side length $L_s$ of 18 $\mu$m to achieve strong evanescent coupling with adjacent resonators.  The coupling gap $g$ between adjacent rings was fixed at $g= 250$ nm.  The corners of the square resonators were rounded with $90^{\circ}$ arcs of $R= 5$ $\mu$m bending radius to reduce bending loss.  The total circumference of each microring was 103 $\mu$m, giving an FSR of about 6.2 nm at 1550 nm wavelength. An SEM image of a unit cell is shown in Fig. \ref{fig:SEMnlongscan} (a).  To excite an edge mode along the bottom lattice boundary, an input waveguide was coupled to a microring at the bottom left corner and the transmitted optical power was measured through an output waveguide coupled to a microring at the bottom right corner.  The input and output couplers were identical to the coupling junctions between microrings. An optical image of the lattice is shown in Fig. \ref{fig:SEMnlongscan} (b).\par

To measure the edge mode spectrum, light from a wavelength-tunable laser (Santec TSL-510) was first passed through a fiber polarizer to obtain TE polarization and then butt-coupled to the input waveguide via a lensed fiber.  The transmitted light in the output waveguide was collected by another lensed fiber and detected by an InGaAs photodetector. Coupling loss between the lensed fiber and the silicon waveguide was estimated to be about 7.5 dB, giving a total input and output coupling loss of 15 dB.  This loss was subtracted from the measured transmitted power and the result normalized by the laser input power (5 mW) to obtain the transmission spectrum of the FLI lattice. 
We also measured the power at the through port of the input waveguide to obtain the reflection spectrum of the lattice.  Figure \ref{fig:SEMnlongscan} (c) shows the transmission spectra at the output port and through port scanned over a broad wavelength range from 1510 nm to 1590 nm with 25 pm increment.  We observe a region of high and relative flat transmission between 1520 nm and 1540 nm, but sharp dips begin to appear for wavelengths longer than 1540 nm.  We attribute this variation in the transmission spectrum to the frequency dispersion of the coupling junctions between microrings.  The upper panel of Fig. \ref{fig:SEMnlongscan} (c) shows the simulated power coupling coefficient of the coupling junction obtained using Ansys Lumerical sofware \cite{Lumerical}. The coupling coefficient is seen to reach near unity over the 1520 - 1540 nm wavelength range, which corresponds to the region of high and flat transmission of the FLI. 
For wavelengths longer than 1540 nm, the coupling becomes less than unity and bulk modes corresponding to the upper and lower dispersive bands are excited, resulting in transmission dips appearing in each FSR.

\begin{figure}[t]
\centering\includegraphics[width=0.8\linewidth]{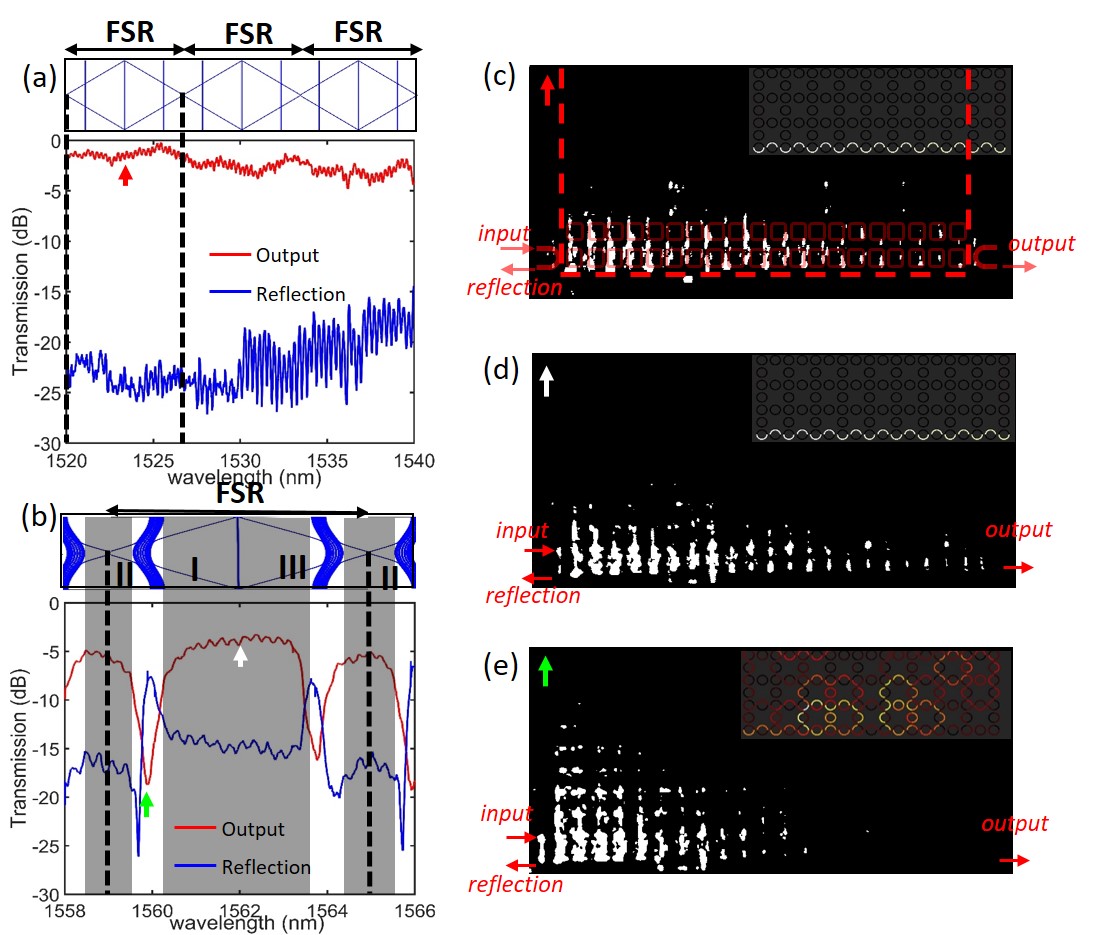}
 \caption{(a), (b) Transmission and reflection spectra of the FLI lattice in the region of near perfect coupling (a) and imperfection coupling (b) (yellow and gray regions in Fig. \ref{fig:SEMnlongscan}(c)).  The band diagram of the lattice is shown above the plot for each case. (c)-(e) NIR images of the scattered light patterns at the wavelengths indicated by the arrows in plots (a) and (b). The insets show the corresponding simulated intensity distribution for each case.}
\label{fig:exp_image}
\end{figure}

To confirm that only edge modes were excited in the region of near perfect coupling, we performed a fine wavelength scan of 2.5 pm resolution from 1520 nm to 1540 nm while also imaging the scattered light intensity distribution at each sampling wavelength using an NIR camera (Sensor Unlimited SU320M) and a 20$\times$ objective lens.  The transmission spectra at the output and through ports are shown in Fig. \ref{fig:exp_image} (a), from which it can be seen that the high transmission remains constant across more than 3 microring FSRs with very little light reflected at the through port.  From the transmission spectrum we estimated the propagation loss of the edge mode to be about 1 dB/mm, 
which is mainly due to scattering at the coupling junctions and ring corners. We note that if flat-band modes were excited, we would have observed transmission dips with spectral widths of at least 20 pm, as estimated from a roundtrip intrinsic loss of about 0.1 dB in the resonant loop of the flat-band mode.  
The absence of such dips implies that only edge modes were excited with negligible coupling into the flat-band modes.  We also obtained direct evidence of edge mode propagation from NIR imaging of the scattered light intensity, which showed a consistent pattern of light localizing along the bottom boundary of the lattice.  A sample image is shown in Fig. \ref{fig:exp_image} (c) at the wavelength where a flat-band mode would exist, as indicated by the red arrow in Fig. \ref{fig:exp_image} (a).  We observe light traveling from input port to output port along the bottom edge, in agreement with the simulated field distribution of the edge mode shown in the inset of the figure.

\begin{figure}[t]
\centering\includegraphics[width=\linewidth]{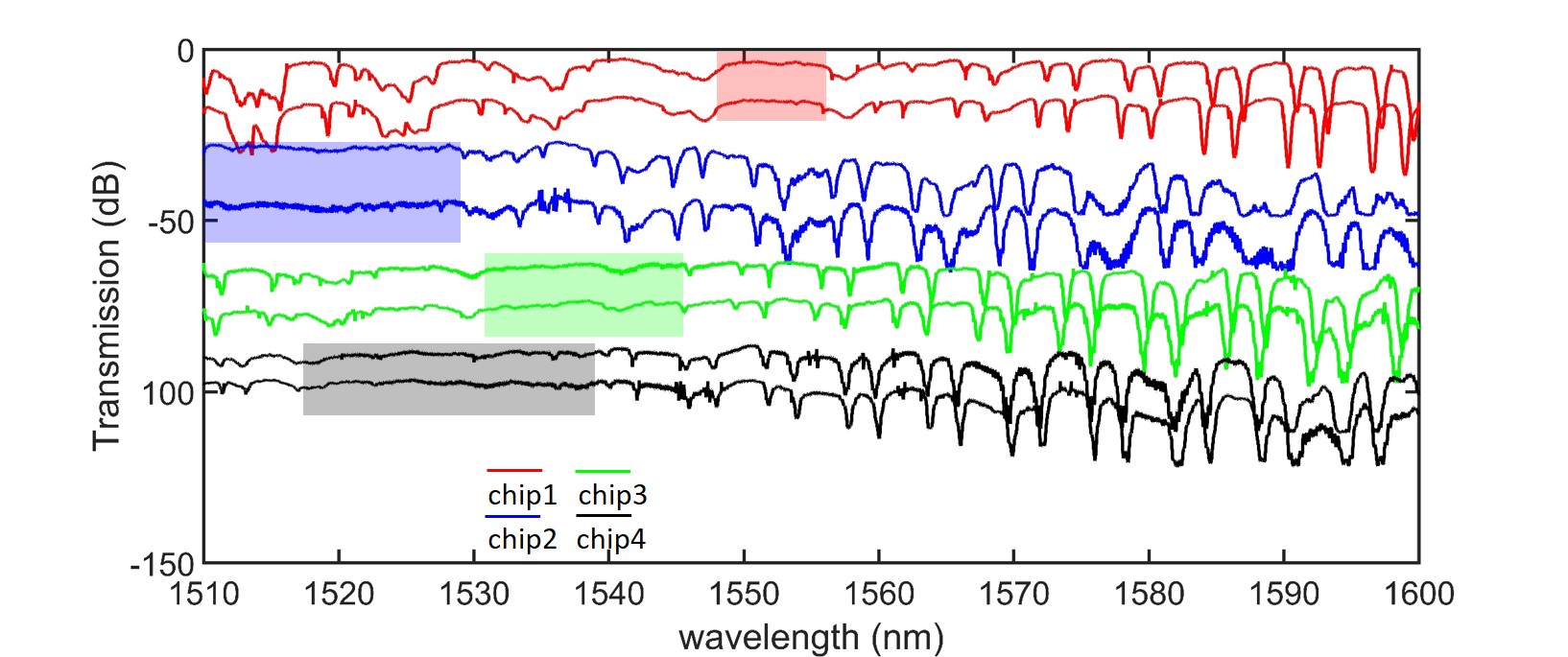}
 \caption{  Transmission measurements of a sample of 8 FLI lattices fabricted on 4 different chips. Each chip has 2 lattices, which are shown by the same color. The traces are shifted in transmission level for clarity.  The shaded regions indicate the 3dB transmission bandwidths of the edge mode continuum obtained when the coupling is near perfect. }
\label{fig:robustnessFLI}
\end{figure}

To contrast the above transmission behavior with the case of imperfect coupling, we show in Fig. \ref{fig:exp_image} (b) a fine wavelength scan of the transmission and reflection spectra from 1558 nm to 1566 nm at 2.5 pm resolution.
Over one FSR, we can identify two high transmission bands (shaded) due to edge mode propagation separated by two regions of strong reflection and attenuation, in agreement with the band diagram of the FLI lattice shown in the upper panel of the plot.
Notably, the transmission band centered around 1562 nm is twice as wide as the transmission bands on the two sides. This indicates that the flat-band mode at zero quasi-energy is not excited since this band remains flat even when the coupling is imperfect. As a result, band gaps I and III are merged together to provide a wide edge mode spectrum.  On the other hand, the upper and lower dispersive bands become broadened when the coupling decreases, so that bulk modes can be easily excited.  These modes spread into the lattice bulk and get partially reflected back into the through port, resulting in the two dips observed in the transmission spectrum and corresponding peaks in the reflection spectrum.  
We also verified the excitation of edge modes and bulk modes by NIR imaging of the scattered light patterns.  
Fig. \ref{fig:exp_image} (d) shows the image obtained at the center wavelength of the broad transmission band where the flat band is located (white arrow in Fig. \ref{fig:exp_image} (b)).  We observe light localized  along the bottom boundary of the lattice forming an edge mode, similar to the image in Fig. \ref{fig:exp_image} (c).  However, when the wavelength was tuned to the middle of a bulk band at 1559.9 nm (green arrow in Fig. \ref{fig:exp_image} (b)), the NIR image in Fig. \ref{fig:exp_image} (e) shows light spreading out into the bulk of the lattice and reflecting back into the input waveguide, with virtually no light reaching the output port. The imaged light patterns are thus also in agreement with the band diagram of the FLI  for the case of imperfect coupling.
We also investigated the robustness of the edge mode continuum to fabrication-induced disorders in both the microring resonant frequencies and couplings by measuring a sample of 8 FLI lattices with identical design parameters fabricated on 4 different chips.
The transmission spectra are shown in Fig. \ref{fig:robustnessFLI}, where traces of the same colors are for different devices on the same chip.  For every trace, we can identify a region of high and flat transmission associated with near perfect coupling, although this region occurs at different wavelengths for lattices on different chips.  We attribute this shift in the transmission band to the systematic variations in the waveguide dimensions from chip to chip.  However, the appearance of a wide edge mode continuum in every lattice, with 3dB bandwidth exceeding a microring FSR, indicates that the flat-band modes are still well localized in energy in the presence of fabrication-induced disorder, with negligible coupling to the edge modes.  We note that the edge mode transmission in the FLI is susceptible only to disorder in the couplings and in the region of near perfect coupling, the coupling coefficient is least sensitive to variations in the coupling angle.  This accounts for the robustness of the edge mode spectrum under near perfect coupling condition.

\section{Conclusion}
We proposed and demonstrated a Floquet-Lieb TPI which combines the extreme frequency localization of flat-band modes and the periodic band structure of Floquet systems to yield an edge mode supercontinuum that can exceed a FB zone.  In addition, the edge modes are shown to be completely immune to disorder in the on-site potential of the lattice.  
We demonstrated a photonic realization of the FLI using 2D coupled microring lattice on an integrated silicon photonic platform.  Under near perfect coupling condition, we achieved a broad edge mode supercontinuum spanning more than 3 FB zones, as verified by both transmission measurements and NIR imaging of the scattered light intensity distributions.  Statistical measurements of random lattices also showed that the edge mode continuum is tolerant to random disorders due to fabrication imperfections. 

The FLI microring lattice provides a promising platform for realizing broadband topological photonic devices with superior robustness.  One potential application of great interest is the generation of topologically-protected entangled photon pairs over a broad frequency range for quantum photonic applications \cite{afzal2023photonic,dai2022topologically}. 
Another application of the FLI is as a topological platform for implementing robust programmable photonic circuits. 
Programmable photonic networks based on recirculating waveguide meshes typically exhibit transmission dips due to the ring resonances, which limits the device bandwidth to less than the ring FSR \cite{song2022robust,zand2020effects,bogaerts2020programmable}.  Our FLI lattice removes this bandwidth limitation while providing robust light transport by the edge modes.

\section*{Appendix A: Derivation of the Hamiltonian of the Floquet-Lieb microring lattice}
We consider a unit cell of the FLI microring lattice as shown in Fig. 1(b).  
The evolution of light over each period can be decomposed into a sequence of 4 coupling steps as shown in the figure.  We define the fields in rings $A$, $B$, and $C$ of unit cell $(m, n)$ as $|\psi_{m,n}(z)\rangle= [\psi^A_{m,n},\psi^B_{m,n},\psi^C_{m,n}]^\text{T}$, with $z$ denoting the direction of light propagation in each ring.  The evolution of the fields over one period can then be described by the Coupled Mode Equations 
\begin{align}
\label{coupled_mode_eqs}
\notag
&-i \frac{\partial \psi_{m,n}^{A}}{\partial z}=\beta \psi_{m,n}^{A}+k_c(1) \psi_{m,n}^{B}+k_c(2) \psi_{m, m}^{C}+k_c(3) \psi_{m-1, n}^{B}+k_c(4) \psi_{m,n-1}^{C}\\
\notag
& -i \frac{\partial \psi_{m,n}^{B}}{\partial z}=\beta \psi_{m,n}^{B}+k_c(1) \psi_{m,n}^{A}+k_c(3) \psi_{m+1, n}^{A} \\
& -i \frac{\partial \psi_{m,n}^{C}}{\partial z}=\beta \psi_{m,n}^{C}+k_c(2) \psi_{m,n}^{A}+k_c(4) \psi_{m,n+1}^{A} \tag{A1}
\end{align}
where $\beta$ is the propagation constant of the microring waveguides and $k_c (j) = 4\theta/L$ is the coupling coefficient in step $j$ (otherwise it is 0). Since the lattice is periodic in $x$ and $y$, we apply Bloch's theorem to express the fields in neighbor unit cells $(m\pm1,n\pm1)$ as $|\psi_{m \pm 1, n\pm 1}\rangle=|\psi_{m, n}\rangle e^{\pm i k_{x}}e^{\pm i k_{y}}$, where we have assumed unit lattice constant.
Applying these boundary conditions to Eq.(\ref{coupled_mode_eqs}), we can express the evolution of the lattice as Eq.(1). 
The Floquet-Bloch Hamiltonian $H_{FB}$ is given by the Hamiltonian $H_j$ of each coupling step,
\cite{zimmerling2022broadband,afzal2018topological}
\begin{equation}
H_{FB}(\textbf{k},z)= \begin{cases}H_1(\textbf{k}), & 0\leq z < L/4 \\ H_2(\textbf{k}), & L/4\leq z < L/2 \\ H_3(\textbf{k}), & L/2\leq z < 3L/4\\ H_4(\textbf{k}), & 3L/4\leq z < L\end{cases}
\label{hamiltonian}
\tag{A4}
\end{equation}
where
\begin{equation}
H_1(\textbf{k})=
  \begin{bmatrix}
         0 & k_c & 0 \\
      k_c & 0 & 0\\
      0 & 0 & 0\\
  \end{bmatrix},
 H_2(\textbf{k})=
  \begin{bmatrix}
         0 & 0 & k_c \\
      0 & 0 & 0\\
      k_c & 0 & 0\\
  \end{bmatrix},
  \notag
\end{equation}

\begin{equation}
      H_3(\textbf{k})=
  \begin{bmatrix}
        0 & k_c e^{-ik_x} & 0 \\
      k_c e^{ik_x} & 0 &0\\
      0 & 0 & 0\\
  \end{bmatrix},
 H_4(\textbf{k})=
  \begin{bmatrix}
       0 & 0 & k_c e^{-ik_y}\\
      0 & 0 & 0\\
      k_c e^{ik_y} & 0 & 0\\
  \end{bmatrix}.
    \notag
\end{equation}

Starting from an initial state $|\psi(\textbf{k},0)\rangle$ at $z = 0$, the evolved state after one period (at $z = L$) is given by $|\psi(\textbf{k},L)\rangle = U_F(\textbf{k})|\psi(\textbf{k},0)\rangle$, where $U_F$ is the Floquet operator,
\begin{equation}
U_F(\textbf{k}) = e^{i H_4(\textbf{k}) L / 4} e^{i H_3(\textbf{k}) L / 4} e^{i H_2(\textbf{k}) L / 4} e^{i H_1(\textbf{k}) L / 4}
\label{Floquet_operator}
\tag{A5}
\end{equation}
Defining $U_j = e^{i H_j L/4}$ as the evolution operator of step $j$, we can compute the Floquet operator as $U_F = U_4 U_3 U_2 U_1$, where
\begin{equation}
    U_1 = \begin{bmatrix}
        \cos \theta & i \sin \theta & 0 \\
        i \sin \theta & \cos \theta & 0\\
      0 & 0 & 1\\
  \end{bmatrix},
      U_2= \begin{bmatrix}
      \cos \theta & 0 & i \sin \theta  \\
       0  & 1 & 0\\
      i \sin \theta & 0 & \cos \theta\\
  \end{bmatrix},
  \notag
\end{equation}

\begin{equation}
    U_3= \begin{bmatrix}
      \cos \theta & i \sin \theta e^{-ik_x} & 0 \\
       i \sin \theta e^{ik_x} & \cos \theta & 0\\
       0 & 0 & 1\\
  \end{bmatrix},
      U_4=\begin{bmatrix}
        \cos \theta & 0 & i \sin \theta e^{-i k_y}  \\
       0  & 1 & 0\\
      i \sin \theta e^{i k_y} & 0 & \cos \theta\\
  \end{bmatrix}.
  \tag{A6}
\end{equation}
From the Floquet operator, we can calculate the effective Hamiltonian as $H_{\mathrm{eff}}= W \Lambda_{\varepsilon} W^{\dagger}$, where $W$ is a matrix containing the eigenvectors of $U_F$ and $\Lambda_{\varepsilon}$ is a diagonal matrix containing the quasi-energies.  For the FLI lattice with perfect coupling, the effective Hamiltonian is explicitly given by Eq.(2).

\section*{Appendix B: Simulation of edge mode transmission spectrum}
\begin{figure}[t]
\centering\includegraphics[width=0.8\linewidth]{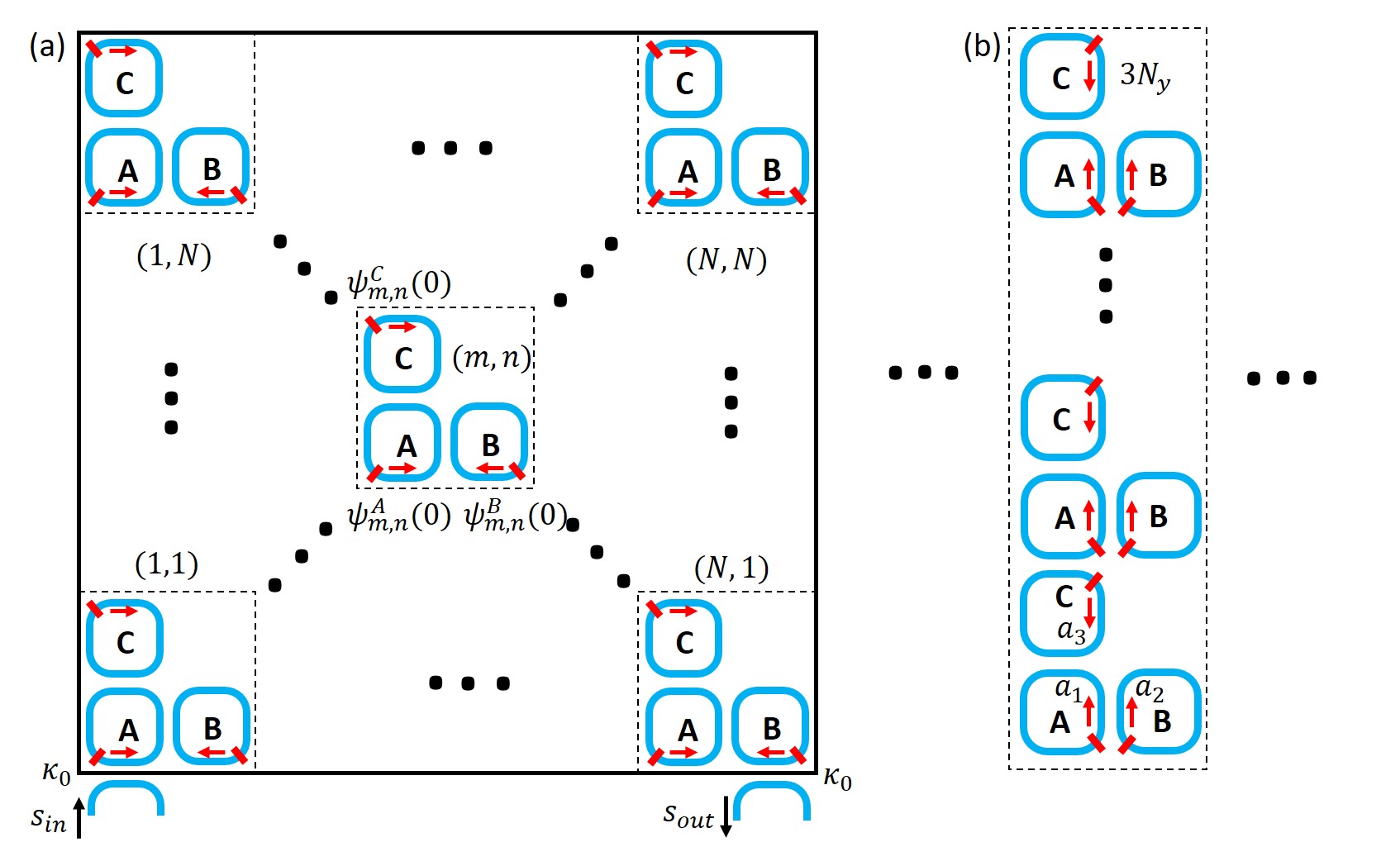}
 \caption{ (a) Schematic of a FLI microring lattice with $N \times N$ unit cells.  The small red bars on each microring indicate the initial position $z = 0$ for light propagation around the microring.  To excite an edge mode along the bottom boundary, an input field $s_{\mathrm{in}}$ is applied to an input waveguide coupled to microring $A$ of cell (1,1) at the bottom left corner.  The transmitted light $s_{\mathrm{out}}$ is obtained through an output waveguide coupled to microring $B$ of cell $(N,1)$ at the bottom right corner. (b) Schematic of a strip of FLI lattice with $N_y$ unit cells in the $y$ direction and infinite extent along $x$.  The microrings are numbered from 1 to $3N_y$ as shown. The small red bars on each microring indicate the initial position $z = 0$ for light propagation around the microring.
 }
\label{fig:Mmatrix}
\end{figure}

We use the power coupling method for coupled microring networks in \cite{van2016optical} to simulate the transmission of an edge mode in an FLI microring lattice. We consider a finite lattice of $N \times N$ unit cells with an input waveguide coupled to microring $A$ of the bottom left unit cell and an output waveguide coupled to microring $B$ of the bottom right unit cell, as shown in Fig. \ref{fig:Mmatrix}.  We number the unit cells as $(m, n)$, $(m, n = 1...N)$, with cell (1,1) at the bottom left corner and cell $(N,N)$ at the top right corner.  
For unit cell $(m, n)$ in the lattice, we denote the fields at position $z$ in microring waveguides $A$, $B$, and $C$ as $[\psi_{m,n}^A(z), \psi_{m,n}^B(z), \psi_{m,n}^C(z)]^{\mathrm{T}} = [a_k(z), a_{k+1}(z), a_{k+2}(z)]^{\mathrm{T}}$, where $k = 3[(n - 1)N + m - 1] + 1$. 
Letting $\textbf{a}(0) = [a_1(0), a_2(0), ...,a_{3N^2}(0)]^{\mathrm{T}}$ be the vector containing the fields in all the microrings at the initial position $z = 0$ (indicated by the red bars in Fig. \ref{fig:Mmatrix}), the fields after propagating one roundtrip are given by $\textbf{a}(L) = \textbf{M} \textbf{a}(0)$, where
\begin{equation}
    \textbf{M}=\textbf{M}_4\textbf{P}^{-1/4} \textbf{M}_3 \textbf{P}^{-1/4}\textbf{M}_2\textbf{P}^{-1/4} \textbf{M}_1\textbf{P}^{-1/4}  
      \tag{A8}
\end{equation}
In the above expression, $\textbf{M}_j$ is the coupling matrix for step $j$ and $\textbf{P} = e^{-\alpha L/2}\mathrm{diag}[e^{i\phi_1}, ..., e^{i\phi_N}]$ is the propagation matrix, with $\phi_k$ being the roundtrip phase of microring $k$ and $\alpha$ the waveguide loss coefficient.  The coupling matrices $\textbf{M}_j$ are constructed as follows.  Starting with $\textbf{M}_j = \textbf{I}$ being a $3N^2 \times 3N^2$ identity matrix, if there is coupling between microring waveguides $k$ and $l$ in step $j$, then $\textbf{M}_j(k,k) = \textbf{M}_j(l,l) = \cos \theta $ and $\textbf{M}_j(k,l) = \textbf{M}_j(l,k) = i\sin \theta$, where $\theta$ is the coupling angle of the FLI lattice. 

For a field $s_{\text{in}}$ applied to the input port, the fields $\textbf{a}(0)$ in the microrings satisfy the equation \cite{van2016optical}
\begin{equation}
    (\textbf{I} - \textbf{L}\textbf{M})\textbf{a}(0) = \textbf{s}
    \tag{A9}
\end{equation}
where $\textbf{s}=[i\kappa_{o} s_{\text{in}}, 0, ..., 0]^{\mathrm{T}}$ and $\textbf{L}$ is a diagonal matrix with diagonal elements
\begin{equation}
\mathbf{L}(k, k)=\left\{\begin{array}{l}
\tau_0, k \in \{2, 3N(N-1)+2\} \\
1, \text { otherwise }
\end{array}\right.
\tag{A10}
\end{equation}
Here $\kappa_{o} = \sin \theta $ and $\tau_{o} = \cos \theta $ are the coupling and transmission coefficients, respectively, of the coupling junction between the input/output waveguides and the microring.  Solving the above equation for $\textbf{a}$ we obtain the power transmission at the output port of the lattice as
\begin{equation}
\label{eq_drop_port}
    T_{\mathrm{out}} = |s_{\mathrm{out}}/s_{\mathrm{in}}|^2 = \kappa_{o}^2|\psi^B_{N,1}(0)|^2/\tau_{o}^2
    \tag{A11}
\end{equation}

To simulate the transmission of the FLI lattice in the presence of randomly distributed microring roundtrip phases, we set the roundtrip phase in microring $k$ to be $\phi_k = \phi_{k}^{\mathrm{nom}} + \delta \phi_k$, where $\phi_{k}^{\mathrm{nom}}$ is the nominal roundtrip phase without perturbation and $\delta \phi_k$ is a random phase error.  Similarly, for random variations in the coupling angles, we add a random term $\delta \theta_k$ to the nominal value of each coupling angle to obtain $ \theta_k = \theta_k^{\mathrm{nom}} + \delta \theta_k$.  The transmission of the lattice is then calculated as described above.

\section*{Appendix C: Orthogonality between flat-band mode and edge mode}
In this section we analytically determine the eigenstates of the edge mode and flat-band mode of a perfectly coupled FLI lattice and show that they are orthogonal at the degenerate frequencies.  We consider a strip of FLI lattice with $N_y$ unit cells in the $y$ direction and infinite extent in the $x$ direction.  We label the microrings $i = 1 ... 3N_y$ as shown in Fig. \ref{fig:strip} and define the state vector containing the fields in the microrings as $|\psi(z)\rangle = [a_1(z), a_2(z), ... a_{3N_y}(z)]^T$.  For the strip, we can explicitly write down the Hamiltonians $H_{sj}$ ($j = 1...4$) of the 4 coupling steps in each period in terms of the unit-cell Hamiltonians $H_1, H_2$ and $H_3$ in Eq.(S4) as
\begin{equation}
    H_{s1}=\operatorname{diag}[\underbrace{H_1, \cdots, H_1}_{N_y}],
        H_{s2}=\operatorname{diag}[\underbrace{H_2, \cdots, H_2}_{N_y}], 
 \notag
\end{equation}

\begin{equation}
        H_{s3}(k_x) =\operatorname{diag}[\underbrace{H_3(k_x), \cdots, H_3(k_x)}_{N_y}], 
         H_{s4}=\operatorname{diag}[\underbrace{\mathbf{0_{2\times 2}}, H_1, \cdots, H_1,\mathbf{0}}_{N_y}]
         \label{strip_Ham}
         \tag{A12}
\end{equation}
where $\mathbf{0_{2\times 2}}$ is the $2\times2$ zero matrix. The Floquet operator of the strip under perfect coupling condition is
\begin{equation}
    U_{sF}= e^{jH_{s4}L/4}e^{jH_{s3}L/4}e^{jH_{s2}L/4}e^{jH_{s1}L/4}
    = \operatorname{diag}[\underbrace{\rho_1(k_x), U_F(k_x),\cdots, U_F(k_x),\rho_2(k_x)}_{N_y}]
    \label{strip_FL}
    \tag{A13}
\end{equation}
where 
\begin{equation}
    \rho_1=\left[\begin{array}{ccc} -e^{-ik_x} & 0 \\ 0 & 0 \end{array}\right], 
    U_F(k_x)=\left[\begin{array}{ccc} 0 & -e^{i k_x} & 0 \\ 0 & 0 & -i e^{-i k_x} \\ -i & 0 & 0\end{array}\right],
    \rho_2=\left[\begin{array}{ccc} 0 & -e^{i k_x}  \\ -1 & 0 \end{array}\right],
    \tag{A14}
\end{equation}
Solving for the eigenvalues of $U_{sF}$, we find that in addition to the flat-band quasi-energies $\varepsilon =\{0,\pm 2\pi/3L\}$, there exist two sets of eigenvalues linear in $k_x$, $\varepsilon=\{ k_x/L, -k_x/2L\}$, which correspond to the two counter-propagating edge states.  The forward-propagating mode ($\varepsilon = k_x/L$) propagates along the bottom lattice boundary while the backward-propagating mode ($\varepsilon = -k_x/2L$) travels along the top boundary (their quasi-energies are different because the two boundaries are not the same).
At the points of degeneracy between the flat-band and edge modes ($\varepsilon L=\{0,\pm 2\pi/3\}$), we can solve for the eigenvectors of these modes (at $z = 0$) to get
\begin{equation}
    |\psi_{\mathrm{fb}}(0)\rangle= [\underbrace{0,i e^{i\varepsilon L},-i e^{-ik_x }e^{-i\varepsilon L},1,0,\cdots,0}_{3N_y}]^T 
    \tag{A15}
\end{equation}
for the flat-band mode and 
\begin{equation}
    |\psi_{\mathrm{edge}}(0)\rangle= [\underbrace{1,0,\cdots,0}_{3N_y}]^T
    \tag{A16}
\end{equation}
for the forward-propagating edge mode.  It can be seen that the two eigenvectors are orthogonal, $\langle\psi_{\mathrm{fb}}(0)|\psi_{\mathrm{edge}}(0)\rangle = 0$.  Furthermore, since the evolution of these modes is given by
\begin{align*}
    |\psi_{\mathrm{fb}}(z)\rangle & = U_s(z) |\psi_{\mathrm{fb}}(0)\rangle \\
    |\psi_{\mathrm{edge}}(z)\rangle & = U_s(z) |\psi_{\mathrm{edge}}(0)\rangle
\end{align*}
where $U_s(z)$ is the unitary evolution operator of the strip, we also have
\begin{equation}
    \langle\psi_{\mathrm{fb}}(z)|\psi_{\mathrm{edge}}(z)\rangle = \langle\psi_{\mathrm{fb}}(0)| U_s^{\dagger}(z) U_s(z) |\psi_{\mathrm{edge}}(0)\rangle = 0
    \tag{A17}
\end{equation}
Thus orthogonality between the flat-band mode and edge mode is maintained over the entire evolution period.

\section*{Appendix D: Immunity of the flat bands to on-site potential disorder}
In this section we show that the flat bands of a FLI with perfect coupling remain flat in the presence of disorder in the diagonal elements of the effective Hamiltonian. The proof involves showing that the eigenvalues of the perturbed Hamiltonian remain independent of the crystal momentum $\textbf{k} = (k_x,k_y)$.  For an infinite FLI lattice with perfect coupling, the flat-band quasi-energies of the effective Hamiltonian are $\varepsilon = \{0, \pm 2\pi/3L\}$, independent of $\textbf{k}$. The effective Hamiltonian of a unit cell can be diagonalized as 
$\textbf{H}_{\mathrm{eff}}(\textbf{k}) = \textbf{W} \mathbf{\Lambda}_{\varepsilon} \textbf{W}^{\dagger}$
where $\mathbf{\Lambda}_{\varepsilon} = \mathrm{diag}[-2\pi/3L, 0, 2\pi/3L]$ and $\textbf{W}$ is the matrix containing the corresponding eigenvectors,
\begin{equation}
\textbf{W}=\frac{1}{\sqrt{3}}\left[\begin{array}{ccc}
i \gamma_y^* & i \gamma_y^* & i \gamma_y^* \\
-e^{-i \phi} & -1 & -e^{i \phi}\\
e^{i \phi} \gamma_x^* & \gamma_x^* & e^{-i \phi} \gamma_x^*
\end{array}\right]
\label{eigenvectors}
\tag{A18}
\end{equation}
where $\gamma_x= e^{ik_x}$, $\gamma_y= e^{ik_y}$ and $\phi = 2\pi/3$.
For a supercell comprised of $N \times N$ unit cells in an infinite unperturbed lattice, the effective Hamiltonian $\textbf{H}_S$ can likewise be diagonalized as
\begin{equation}
    \textbf{H}_S(\textbf{k}) = \textbf{V} \mathbf{\Lambda}_S \textbf{V}^{\dagger}
    \label{unperturbed_system}
    \tag{A19}
\end{equation}
where $\mathbf{\Lambda}_S = \mathrm{diag}[\mathbf{\Lambda}_{\varepsilon},\cdots,\mathbf{\Lambda}_{\varepsilon}]$ and $\textbf{V} = \mathrm{diag}[\mathbf{W},\cdots,\mathbf{W}e^{i(mk_x+nk_y)},\cdots\mathbf{W}e^{i(Nk_x+Nk_y)}]$ are block diagonal matrices of size $3N^2 \times 3N^2$ (the indices $m$ and $n$ indicate unit cell $(m,n)$).

\begin{figure}[t]
\centering\includegraphics[width=0.8\linewidth]{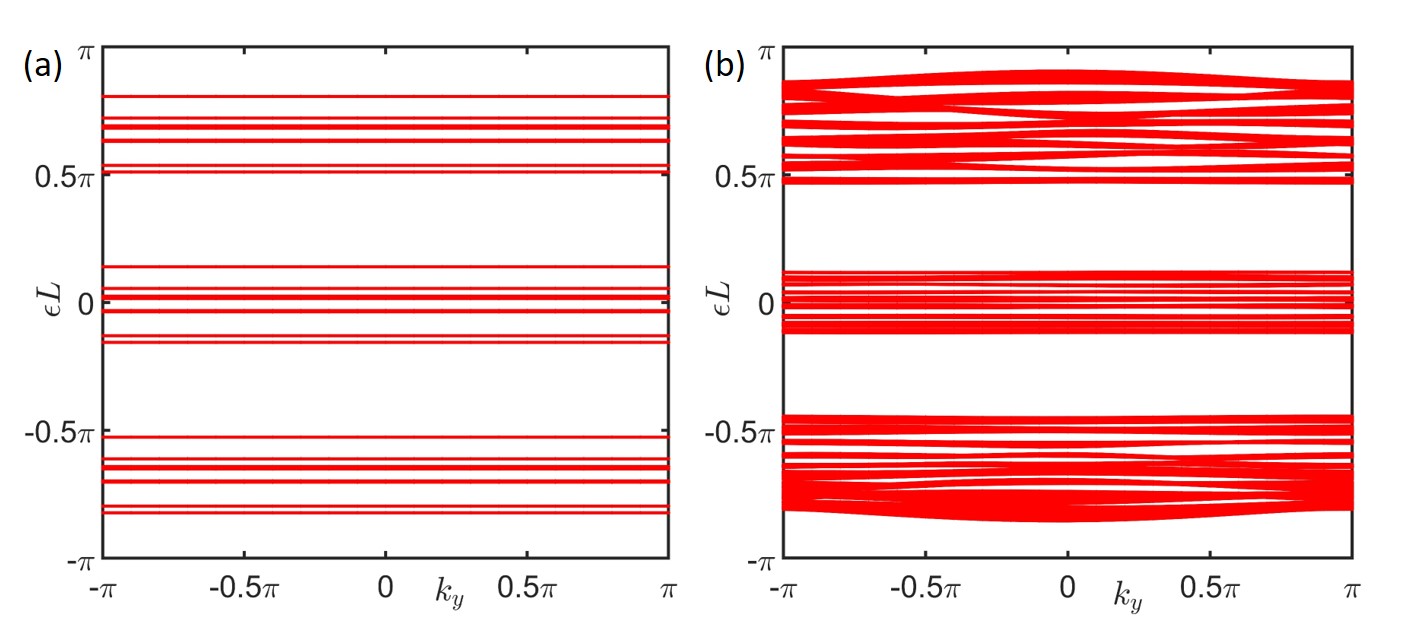}
 \caption{ Numerical verification of the robustness of the flat bands of an FLI against random diagonal disorder for (a) perfect coupling and (b) imperfect coupling ($\theta=0.4\pi$). We applied random phase disorder $\delta \phi_i \in [-\pi/4,\pi/4]$ to each microring and calculate the projected band diagram of a $3\times 3$ super cell.}
\label{fig:disorderdiagonal}
\end{figure}

Suppose we now introduce random disorder terms $\delta_i$ ($i = 1 \cdots 3N^2)$ to the diagonal elements of the Hamiltonian of the supercell.  The Hamiltonian of the perturbed system is given by $\widetilde{\textbf{H}}_S = \textbf{H}_S + \textbf{D}$, where $\textbf{D} = \mathrm{diag}[\delta_1, \cdots, \delta_{3N^2}]$.  We diagonalize $\widetilde{\textbf{H}}_S$ as
\begin{equation}
    \widetilde{\textbf{H}}_S = \textbf{H}_S + \textbf{D} = \widetilde{\textbf{V}} \widetilde{\mathbf{\Lambda}}_S \widetilde{\textbf{V}}^{\dagger}
    \label{perturbed_system}
    \tag{A20}
\end{equation}
where $\widetilde{\mathbf{\Lambda}}_S$ is a diagonal matrix containing the perturbed eigenvalues with associated eigenvector matrix $\widetilde{\textbf{V}}$.  By expressing the perturbed eigenvectors $|\widetilde{\psi}_n\rangle$ in the basis of the unperturbed Bloch modes $|\psi_n\rangle$  (the column vectors in $\textbf{V}$)
\begin{equation}
    |\widetilde{\psi}_n\rangle=\sum_{m=1}^{3 N^2} T_{n, m} |\psi_m\rangle, (n = 1 \cdots 3N^2)
    \tag{A21}
\label{basiseq}
\end{equation}
we obtain $\widetilde{\textbf{V}} = \textbf{VT}$, where $\textbf{T}$ is an $3N^2 \times 3N^2$ matrix with elements $T_{m,n}$.  Using this in Eq.(\ref{perturbed_system}), we get
\begin{equation}
    \widetilde{\textbf{H}}_S = \textbf{H}_S + \textbf{D} = \textbf{VT} \widetilde{\mathbf{\Lambda}}_S \textbf{T}^{\dagger} \textbf{V}^{\dagger}
    \tag{A22}
\end{equation}
or
\begin{equation}
    \textbf{V}^{\dagger} \widetilde{\textbf{H}}_S \textbf{V} = \textbf{V}^{\dagger}(\textbf{H}_S + \textbf{D})\textbf{V} = \textbf{T}\widetilde{\mathbf{\Lambda}}_S \textbf{T}^{\dagger}
    \tag{A23}
\end{equation}
With the use of Eq.(\ref{unperturbed_system}), we obtain
\begin{equation}
    \textbf{V}^{\dagger}\widetilde{\textbf{H}}_S \textbf{V} = \mathbf{\Lambda}_S + \textbf{V}^{\dagger}\textbf{D}\textbf{V} = \textbf{T}\widetilde{\mathbf{\Lambda}}_S \textbf{T}^{\dagger}
    \tag{A24}
\end{equation}
which shows that the perturbed Hamiltonian $\widetilde{\textbf{H}}_S$ and $\mathbf{\Lambda}_S + \textbf{V}^{\dagger}\textbf{D}\textbf{V}$ are similar matrices with the same eigenvalues (matrix $\widetilde{\mathbf{\Lambda}}_S$). Since $\mathbf{D}$ is diagonal and $\mathbf{V}$ is block diagonal, the matrix $\textbf{V}^{\dagger}\textbf{D}\textbf{V}$ also has a block diagonal form,
\begin{equation}
    \textbf{V}^{\dagger}\textbf{D}\textbf{V}=\mathrm{diag}[\mathbf{E}_1,\mathbf{E}_2,\cdots,\mathbf{E}_{N^2}]
    \tag{A25}
\end{equation}
where $\mathbf{E}_n= \mathbf{W}^{\dagger} \mathbf{D}_n \mathbf{W}$, and $\mathbf{D}_n$ is a diagonal matrix containing the on-site perturbations $(\delta_j, \delta_k, \delta_l)$ in unit cell $n$,
$\mathbf{D}_n= \mathrm{diag}[\delta_j,\delta_k,\delta_l]$.  With $\mathbf{W}$ given in Eq.(\ref{eigenvectors}), we can explicitly compute $\mathbf{E}_n$ to get
\begin{equation}
\begin{aligned}
\mathbf{E}_n =\frac{1}{3}\left[\begin{array}{ccc}
\delta_j+\delta_k+\delta_l & \delta_j +\delta_k e^{i \phi}+\delta_l e^{-i \phi} & \delta_j +\delta_k e^{2 i \phi} +\delta_l e^{-2 i \phi} \\
\delta_j +\delta_k e^{-i \phi}+\delta_l e^{i \phi} & \delta_j+\delta_k+\delta_l & \delta_j +\delta_k e^{i \phi}+\delta_le^{-i \phi} \\
\delta_j +\delta_k e^{-2 i \phi}+\delta_l e^{2 i \phi} & \delta_j +\delta_k e^{-i \phi}+\delta_l e^{i \phi} & \delta_j+\delta_k+\delta_l
\end{array}\right]
\end{aligned}
\tag{A26}
\end{equation}
which is independent of $k_x$ and $k_y$.  The matrix $\textbf{V}^{\dagger}\textbf{D}\textbf{V}$ is thus independent of $\textbf{k}$ and as a result, the eigenvalues of the perturbed Hamiltonian $\widetilde{\textbf{H}}_S$ are also dependent of $\textbf{k}$, implying that the quasi-energy bands remain flat in the presence of perturbations to the on-site potentials.
As a numerical verification, we simulated a supercell comprised of $3 \times 3$ unit cells subject to random disorders in the microring resonance frequencies, or roundtrip phases.  We applied a uniformly-distributed perturbation $\delta \phi_i$ ($-\pi/4 < \delta \phi_i < \pi/4$) to the roundtrip phase of microring $i$. The computed band diagrams are shown in Figs. \ref{fig:disorderdiagonal}(a) and (b) for the lattice with perfect coupling and imperfect coupling, respectively.  It can be seen that the bands of the perfectly-coupled lattice remain flat while those of the lattice with imperfect coupling (including the bands at $\varepsilon_0$) become dispersive in the presence of random variations in the microring resonant frequencies.

\newpage
\begin{backmatter}
\bmsection{Funding}
This work was supported by the Natural Sciences and Engineering Research Council of Canada.
\bmsection{Acknowledgments}
H. Song performed modeling, numerical simulations and data analyses of FLI devices. T. Zimmerling designed the chips. H. Song and T. Zimmerling performed transmission measurements and NIR imaging of the chips. B Leng performed SEM and optical imaging of the devices. H. Song and V. Van wrote the paper, with contributions from all the authors.
\bmsection{Disclosures}




The authors declare no conflicts of interest.

\bmsection{Data Availability Statement}
The data that support the findings in this paper and supplemental document are available from the corresponding author upon reasonable request.


\end{backmatter}

\bibliography{Reference}






\end{document}